\documentclass[10pt]{article}
\usepackage{a4wide}                     % to get a wider page margen
\usepackage{amsfonts,amsmath}            % to make fancy symbols
\usepackage{amssymb,latexsym}            % to make more fancy symbols
\usepackage{epsf}                        % to make figures
\newcommand{\sect}[1]{\setcounter{equation}{0}\section{#1}}

\def\be{\begin{equation}}
\def\ee{\end{equation}}
\def\bea{\begin{eqnarray}}
\def\eea{\end{eqnarray}}
\def\nn{\nonumber \\ [.3cm]}

\def\hsp#1{\hspace{#1}}

\def\part{\partial}
\def\tfrac#1#2{{\textstyle{\frac{#1}{#2}}}}

\def\x{\times}

\def\incl{\mbox{i}}

\def\C{\ensuremath{\mathbb{C}}}

\def\Ortin{Ort{\'\i}n}

\def\makeatletter{\catcode`\@=11}% 11:letter
\makeatletter
\def\mathbox#1{\hbox{$\m@th#1$}}%
\def\math@ccstyles#1#2#3#4#5#6#7{{\leavevmode
      \setbox0\mathbox{#6#7}%
      \setbox2\mathbox{#4#5}%
      \dimen@ #3%
      \baselineskip\z@\lineskiplimit#1\lineskip\z@
      \vbox{\ialign{##\crcr
             \hfil \kern #2\box2 \hfil\crcr
             \noalign{\kern\dimen@}%
             \hfil\box0\hfil\crcr}}}}
\def\mathaccstyles{\math@ccstyles\maxdimen}
\def\maththroughstyles{\math@ccstyles{-\maxdimen}}
\def\unity%
 {\maththroughstyles{.45\ht0}\z@\displaystyle {\mathchar"006C}\displaystyle 1}
\begin{document}

\rightline{\small KUL-TF-2003/05}
\rightline{\small FFUOV-03/02}
\rightline{hep-th/0303183}
\rightline{March 2003}
\vspace{2truecm}

%%%%%%%%%%%%%%%%%
\centerline{\LARGE \bf A Microscopical Description of Giant Gravitons II:}
\vspace{.5truecm}
\centerline{\LARGE \bf The $AdS_5\times S^5$ Background }
\vspace{1.3truecm}

\centerline{
    {\large \bf Bert Janssen${}^{a,}$}\footnote{E-mail address: 
                                  {\tt bert.janssen@fys.kuleuven.ac.be} },
    {\large \bf Yolanda Lozano${}^{b,}$}\footnote{E-mail address: 
                                  {\tt yolanda@string1.ciencias.uniovi.es}}
    {\bf and}
    {\large \bf Diego Rodr\'{\i}guez-G\'omez${}^{b,c,}$}\footnote{E-mail address:
                                  {\tt diego@fisi35.ciencias.uniovi.es}}
                                                            }

\vspace{.4cm}
\centerline{{\it ${}^a$Institute for Theoretical Physics, Katholieke Universiteit Leuven,}}
\centerline{{\it Celestijnenlaan 200D, B-3001 Leuven, Belgium}}

\begin{center}
{\it ${}^b$Departamento de F{\'\i}sica,  Universidad de Oviedo, \\  
          Avda.~Calvo Sotelo 18, 33007 Oviedo, Spain}
\end{center}

\begin{center}
{\it ${}^c$Departamento de F\'{\i}sica Te\'orica, Universidad Aut\'onoma de Madrid,\\
          Cantoblanco, 28049 Madrid, Spain}
\end{center}
\vspace{2truecm}

%%%%%%%%%%%%%%%%%
\centerline{\bf ABSTRACT}
\vspace{.5truecm}

\noindent In this article we continue with the microscopical investigation
of giant graviton configurations in $AdS_m\times S^n$ spacetimes, initiated in
hep-th/0207199. Using dualities and a Matrix theory derivation we propose an
action that describes multiple Type IIB gravitons. This action contains multipole
moment couplings to the Type IIB background potentials. Using these couplings,
we study, from the microscopical point of view, the giant graviton and dual giant
graviton configurations in the $AdS_5\times S^5$ background. In both cases the
gravitons expand into a non-commutative 3-sphere, that is defined as an 
$S^1$-bundle over a fuzzy 2-sphere. When the number of gravitons is large we find perfect
agreement with the Abelian, macroscopical description of giant gravitons in this
spacetime, given in the literature.

%This file is a model of a title page.

%%end of title page
%%%%%%%%%%%%%%%%%%%%%%%%%%%%%%%%%%%%%%%%%%%%%%%%%%%%%%%%%
\newpage
%%%%%%%%%%%%%%%%%%%%%%%%%%%%%%%%%%%%%%%%%%%%%%%%%%%%%%%%%
\sect{Introduction}

Giant gravitons \cite{GST, GMT, HHI,DJM} are stable brane 
configurations with non-zero angular momentum, that are wrapped around
$(n-2)$- or $(m-2)$-spheres
in $AdS_m \x S^n$ spacetimes, and carry a dipole moment with respect to the
background gauge potential. 
They are not topologically stable, but are at a dynamical 
equilibrium because the contraction due to the tension of the brane is precisely 
cancelled by the expansion due to the coupling of the angular momentum to the background 
flux field.  It turns out that these spherical brane configurations are massless, 
conserve the same number of
supersymmetries and carry the same quantum numbers of a graviton. The fact 
that they are extended objects of finite size has lead to the name of giant gravitons.  

These configurations were first proposed in \cite{GST} as a way to satisfy the stringy
exclusion principle implied by the AdS/CFT correspondence. The spherical $(n-2)$-brane
expands into the $S^n$ part of the geometry with a radius proportional to its
angular momentum. Since this radius is bounded by the radius of the $S^n$, the
configuration has associated a maximum angular momentum.
The $(m-2)$-brane giant graviton configurations later found in \cite{GMT,HHI} expand, 
on the other hand,
into the $AdS_m$ component of the spacetime, so that there is no upper bound implied on
their angular momentum. They are referred to in the literature \cite{GMT} as dual
giant gravitons. One expects that quantum mechanically these different states will tunnel
into each other, forming a unique ground state. In \cite{GMT} it is speculated that
the stringy exclusion principle may still be satisfied if  
once the quantum mixing is taken into account there is no supersymmetric
ground state for angular momenta
beyond the exclusion principle bound. Other giant graviton solutions in various backgrounds
have been studied elsewhere in the literature \cite{CR1,CR2}. Giant gravitons
in pp-wave backgrounds have been studied in \cite{BMN,ST,TT,KY,YY}.

The appearance of giant gravitons as blown up massless particles hints to a connection 
with other examples of expanded brane configurations, more precisely to the dielectric effect, 
where multiple coinciding D$p$-branes can expand into higher dimensional D-brane
configurations.  
There are two complementary descriptions of this effect. Consider the case in which the
D$p$-branes expand into a spherical D$(p+2)$-brane.
The first one is an Abelian description, describing 
the spherical D$(p+2)$-brane with a large number of D$p$-branes dissolved on its worldvolume
\cite{roberto}. The second one is a non-Abelian formulation \cite{TvR1,TvR2,Myers},
describing how multiple coinciding
D$p$-branes expand into a D$(p+2)$-brane with the topology of a fuzzy 2-sphere \cite{Myers}. 
Both descriptions agree in the limit 
where the number $N$ of D$p$-branes is very large.

It has been suggested in the literature (see for example 
\cite{DTV,L,BMN,JL1}) that it should be possible to describe the giant
gravitons of \cite{GST,GMT,HHI,DJM} in terms of dielectric gravitational waves.
Since massless particles, in particular gravitons, are the source terms for 
gravitational waves, it is natural to expect that a dielectric effect for these 
gravitational waves will provide a microscopic picture for the giant graviton 
configurations. By analogy with the dielectric effect for D-branes,
it is believed that in the 
limit when the number of gravitons, $N$, is large, this non-Abelian, microscopical 
description should match the Abelian, macroscopical description of \cite{GST,GMT,HHI,DJM},
in terms of an spherical brane with angular momentum, where microscopically
the angular momentum of the Abelian description is
interpreted as the total momentum of the multiple gravitational waves, propagating in
the spherical part of the geometry. 

In \cite{JL1} an action describing Type IIB gravitational waves was derived using 
Matrix string theory in a weakly curved background.
This action shows the linear couplings to closed string backgrounds, 
and contains the now familiar multipole couplings that give
rise to the dielectric effect \cite{Myers}. From them one can construct
configurations of multiple coinciding gravitational waves expanding into
higher dimensional fuzzy surfaces. 
However, with a non-Abelian action for gravitational waves only known
up to linear order in the background fields it is not possible to
check whether dielectric gravitational waves really provide a 
microscopical description for the giant gravitons in $AdS_m\times S^n$
backgrounds, given that the $AdS_m\times S^n$ spacetime cannot
be taken as a linear perturbation to Minkowski.

In reference \cite{JL2} we gave one step further in this direction, by
providing a closed expression for the worldvolume action for multiple
M-theory gravitons, valid beyond the linear approximation. A linear action
for M-theory gravitons can easily be constructed by just uplifting the
linear action for multiple Type IIA gravitons constructed in 
\cite{JL1} using Matrix string theory in a weakly curved 
background\footnote{The derivation in \cite{JL1} considers Matrix string
theory in a weakly curved background in the Sen-Seiberg limit, so that
the strings carry spatial momentum, and then takes static gauge.
In this way one arrives at an action that describes multiple massless
particles carrying momentum. In the Abelian limit, this action is related to
the perturbative action for massless particles through a Legendre transformation
(we refer the reader to \cite{JL1} for the details of this construction).}. 
Demanding consistency with the action for
coincident D0-branes when the gravitons propagate along the eleventh direction
it is possible to extend this action beyond the linear approximation used 
in the Matrix theory calculation. In fact, a non-trivial check of this action
is that it predicts the polarisation of the gravitons into fuzzy surfaces,
in such a way that in $AdS_m\times S^n$ backgrounds the energy and radius of 
the dielectric configurations exactly match those of the giant graviton
solutions of \cite{GST,GMT}\footnote{The explicit calculation in \cite{JL2}
is for the giant graviton solution in $AdS_7\times S^4$ and the dual giant
graviton solution in $AdS_4\times S^7$, i.e. the cases involving fuzzy
2-spheres.}.

With a closed expression for the action for M-theory gravitational waves
it is possible to construct an action for Type IIA gravitational waves valid
beyond the linear approximation of \cite{JL1}. This action was presented
in \cite{JL2}. Given that Type II waves are simply related by a T-duality
transformation, it is straightforward to construct an action for multiple
Type IIB gravitational waves valid beyond the linear approximation of \cite{JL1}.
With this action it is then possible to construct explicit configurations
of multiple Type IIB gravitons in the $AdS_5\times S^5$ background, which we
can compare with the 
giant graviton \cite{GST} and the dual giant graviton \cite{GMT,HHI} solutions
in this spacetime.
This is precisely the aim of this paper.  

The construction of non-Abelian giant gravitons in $AdS_5 \x S^5$ turns out to be more 
involved than the eleven-dimensional non-Abelian giant graviton solutions of \cite{JL2}. In 
this background 
the gravitational waves expand into a spherical D3-brane, and hence a fuzzy 3-sphere 
Ansatz is needed in the non-Abelian construction, rather than the well-known fuzzy 2-sphere,
in terms of 
$SU(2)$ matrices. Non-commutative odd spheres have been constructed in \cite{GR, R1,R2}, as 
subspaces of fuzzy even spheres. It turns out that these constructions are not applicable 
to our case. The main reason is that due to our construction of the action for Type IIB 
gravitational waves via T-duality, this action exhibits an extra isometry direction associated 
to the coordinate over which the T-duality was performed. This isometry will require a 
non-manifestly $SO(4)$-symmetric solution, which is not of the type presented in \cite{GR, R1,R2}.
Furthermore, the fact that we are dealing with gravitational waves, which have a one-dimensional 
worldvolume, allows only non-Abelian couplings similar to the ones one encounters in the case 
of fuzzy two-spheres. The solution to this apparent paradox is to consider the three-sphere as
a $U(1)$-bundle over $S^2$ and choose the extra isometry direction to be the fibre coordinate. 
The fuzzy version of the $S^3$ will then consist of a (Abelian) $U(1)$-fibre over a fuzzy $S^2$.
   
The paper is organised as follows: we start in Section 2 with the construction of the action 
for Type IIB gravitational waves in the way explained in the previous paragraphs.
This action is valid beyond the linear order approximation taken in \cite{JL1}, 
and can therefore be used to study the $AdS_5\times S^5$ background considered 
in this paper. With this action, we provide, in Section 3, a microscopical description 
of the giant graviton configuration of \cite{GST} in $AdS_5\times S^5$. This solution 
is expected to occur, microscopically, in the form of $N$ gravitons polarised in a 
fuzzy 3-sphere contained in $S^5$. Describing the $S^3$ as an $S^1$ bundle over $S^2$ 
and making non-commutative the base 2-sphere we find a giant graviton solution which 
in the large $N$ limit reproduces exactly the (Abelian) result of \cite{GST}. This is 
a non-trivial check of the validity of our microscopical description and, in particular, 
of our non-commutative Ansatz. Further support is provided by the calculation in Section 4. 
In this section we construct microscopically the dual giant graviton
solution of \cite{GMT,HHI}. Describing the 3-sphere contained in $AdS_5$ as an $S^1$ bundle
over $S^2$ and making this $S^2$ non-commutative we find a dual giant graviton solution 
which is also in perfect agreement with the results in \cite{GMT,HHI} when $N$ is large. 
Section 5 contains the Discussion, where we comment on the relations between our non-commutative 
3-sphere solutions and other constructions of odd non-commutative spheres previously discussed in 
the literature \cite{GR,R1,R2}.

%%%%%%%%%%%%%%%%%%%%%%%%%%%%%%%%%%%%%%%%%%%%%%%%%%%%%%%%%%%%%%%%%%%%%%%%%%%%%%%%
\sect{The action for multiple Type IIB gravitational waves}

In this section we construct an action for Type IIB gravitational waves suitable
for the study of giant graviton configurations in the $AdS_5\times S^5$ 
background\footnote{We will be dealing throughout the article with only the
bosonic part of the worldvolume effective actions. This is enough for the study
of the bosonic giant graviton configurations.}.
The starting point is the action for multiple Type IIA gravitational waves presented in
\cite{JL2}. This action is simply the weakly coupled (Type IIA) version of the action 
there proposed for the microscopical study
of M-theory gravitons in $AdS_m\times S^n$ spacetimes. The M-theory action of \cite{JL2}
gives the couplings of multiple gravitational waves to M-theory background fields,
in the form of a closed expression valid beyond the weak background field limit.
The extension beyond the linear order of Matrix theory can be done
by demanding agreement with the action for multiple
D0-branes when the gravitons propagate along the eleventh direction. 
T-dualising the action for the Type IIA waves we will obtain an action for Type IIB
waves also valid beyond the linear order approximation taken in \cite{JL1}.

The Born-Infeld action proposed in \cite{JL2} to describe multiple Type IIA gravitational waves
is given by
\begin{equation}
\label{IIAwavesBI}
S^{\rm BI}_{{\rm W}_A}=-T_0\int d\tau~ {\rm Str}\Bigl\{ k^{-1}
\sqrt{-P[E_{00}+E_{0i}(Q^{-1}-\delta)^i_k
E^{kj}E_{j0}] \ {\rm det} (Q^i_j )}
\ \Bigr\},
\end{equation}
where
\begin{eqnarray}
E_{\mu\nu} &=& g_{\mu\nu}-k^{-2}k_\mu k_\nu
                   + k^{-1}e^\phi (\incl_k C^{(3)})_{\mu\nu},  \\
Q^i_j &=& \delta^i_j + i [X^i,X^k]e^{-\phi}k E_{kj}\, , \qquad i,j=1,\dots 9\, .
\end{eqnarray}
In this action the direction of propagation of the waves occurs as an
isometry direction, $k^\mu$ being the Killing vector pointing in this 
direction\footnote{This is inferred by the analysis of the monopole
term in the Chern-Simons effective action, which shows that the momentum 
arises as the charge with respect to the background field $k^{-2}k_\mu$.
For details see \cite{JL1}.
%$$\int d\tau k^{-2}k_\mu \dot{X}^\mu=\int d\tau (k^{-2}k_a\dot{X}^a+
%\dot{X}^9)\, ,$$
%where $\mu=(a,9)$ and we are taking $k^\mu=\delta^\mu_9$. Therefore,
%$$P_9=\frac{\delta S}{\delta \dot{X}^9}=NT_0\, .$$
}.
In our notation $k^2=g_{\mu\nu}k^\mu k^\nu$ and 
$(\incl_k C^{(3)})_{\mu\nu}= k^\rho C^{(3)}_{\rho\mu\nu}$. 
This action is manifestly invariant under global gauge
transformations along the Killing direction:
$\delta X^\mu=\Lambda k^\mu$, since this direction is projected out through the
effective metric ${\cal G}_{\mu\nu}=g_{\mu\nu}-k^{-2}k_\mu k_\nu$ and the
contraction of the 3-form with $k^\mu$. The condition that $k^\mu {\cal G}_{\mu\nu}=0$
implies that the gravitational field is transversal to the direction of propagation
of the waves. For simplicity, this action has been calculated for
vanishing $C^{(1)}$, $B^{(2)}$ and the worldvolume scalar field $A$.\footnote{The 
worldvolume scalar field comes from the reduction of the eleventh scalar, and forms an
invariant field strength with the RR 1-form potential.} 

We now make a T-duality transformation along a tranverse direction $Z$ in order to 
obtain the action for Type IIB gravitational waves. This transformation was explained 
in detail in reference \cite{JL1}, for the linearised action. The action for Type IIB
gravitational waves contains two worldvolume scalars $A$ and $Z$, being, respectively,
the T-dual 
of the Type IIA worldvolume scalar $A$ and the T-dual of the embedding scalar in the direction
of the dualisation. Their gauge invariant curvatures are given by 
$F = \partial A-P[\incl_l C^{(2)}]$ and ${\tilde F}=\partial Z+P[\incl_l B^{(2)}]$ and 
together they form a doublet under the Type IIB S-duality transformations. Consistently with 
S-duality and the truncation imposed in the Type IIA action, we will set $C^{(2)}$,
$B^{(2)}$, $A$ and $Z$ equal to zero. Also, for simplicity we take $k_z=0$.
This truncation is suitable for the study of gravitational waves in
the $AdS_5\times S^5$ background, as we show below. Hence, we obtain the following 
Born-Infeld action:
\begin{equation}
\label{IIBwavesBI}
S^{\rm BI}_{{\rm W}_B}=-T_0\int d\tau~ {\rm Str}\Bigl\{ k^{-1}
\sqrt{-P[E_{00}+E_{0i}(Q^{-1}-\delta)^i_k
E^{kj}E_{j0}] \ {\rm det} (Q^i_j )}
\ \Bigr\},
\end{equation}
where now
\begin{eqnarray}
\label{EQ}
E_{\mu\nu} &=& g_{\mu\nu}-k^{-2}k_\mu k_\nu -l^{-2}l_\mu l_\nu
-k^{-1}l^{-1} e^{\phi}(\incl_k\incl_l C^{(4)})_{\mu\nu}, \\
Q^i_j &=& \delta^i_j + i[X^i,X^k]e^{-\phi} k l E_{kj}.
\end{eqnarray}
Here $l^\mu$ is a new Killing vector, pointing along the direction in which we
performed the T-duality transformation. It is easy to check that
this action is invariant under the global isometric transformations generated by $k^\mu$
and $l^\mu$: $\delta X^\mu=\Lambda^{(1)}k^\mu+\lambda^{(2)}l^\mu$, since these 
directions are projected out through the effective metric
${\cal G}_{\mu\nu}=g_{\mu\nu}-k^{-2}k_\mu k_\nu -l^{-2}l_\mu l_\nu$ and the 
double contraction of the 4-form with the two Killing vectors. It is due to the fact that we
have two isometric directions that $C^{(4)}$ can couple to the Born-Infeld part
of the action. This coupling plays a key role in the microscopical description of the 
$AdS_5 \times S^5$ giant graviton solution \cite{GST}, that we will perform in the next section.
In this action $k^\mu$ still points in the direction of propagation of the waves, since under
the T-duality transformation and with the truncations above
the monopole term is mapped onto itself. The other isometric
direction, $Z$, is however not physical, but just an artifact of the T-duality transformation.
We are therefore describing waves which are smeared in the $Z$-direction. In the Abelian limit,
(\ref{IIBwavesBI}) is still a complicated expression containing two isometric directions.
However, together with the (Abelian) Chern-Simons coupling, which is given by \cite{JL1}:
\begin{equation}
S_{\rm W_B}^{\rm CS}=T_0\int d\tau k^{-2}k_\mu\partial X^\mu\, ,
\end{equation}
this effective action can be related to the (dimensional reduction along the
$Z$ direction of the) usual perturbative action for massless particles
\begin{equation}
S[\gamma]=-\frac{NT_0}{2}\int d\tau \sqrt{|\gamma|}\gamma^{-1}\partial X^\mu
\partial X^\nu g_{\mu\nu}
\end{equation}
by means of a Legendre transformation that restores the dependence on the
direction of propagation (see section 2 in \cite{JL1} for the details).
In the non-Abelian case, however, it is not possible to restore this dependence.
First of all, this direction is not matrix-valued, so even though we could in
principle restore some explicit dependence on its time derivative, the new terms
would be Abelian. Second, it is clear that the Legendre transformation cannot give
rise to non-Abelian commutators involving this direction, so we would end up in any
case with an action with reduced transverse rotational invariance. Thus, we are constrained
to work, in the non-Abelian case, with an action with two isometries, and assume
the presence of the extra unphysical isometry.

This isometry will however play a key role in our microscopical description of giant
graviton configurations in $AdS_5\times S^5$. Indeed, the presence of this compact
isometry suggests the representation of the 3-sphere of the giant graviton configurations
as a $U(1)$ bundle over $S^2$, with the $U(1)$ invariance being associated to 
translations along this direction. We will see that the giant graviton configurations
correspond to the polarisation of the gravitons in fuzzy 3-spheres represented as $U(1)$
bundles over a fuzzy $S^2$.

Finally, we turn to the Chern-Simons part of the action for the Type IIB waves.
This action was constructed, to linear order in the background fields, in
reference \cite{JL1} (see expression (4.5)). In particular, the linear coupling to 
the 4-form RR-potential, relevant for the construction of giant gravitons in the  
$AdS_5\times S^5$ background, was shown to be given by:
\begin{equation}
\label{IIBwavesCS}
S^{\rm CS}_{\rm W_B}=-i\int d\tau\, {\rm STr} \{P[(\incl_X\incl_X)\incl_l C^{(4)}]\}.
\end{equation}
In this action the pull-backs into the worldvolume are defined in terms of
gauge covariant derivatives
\begin{equation}
{\cal D}X^\mu=\partial X^\mu-A^{(1)}k^\mu-A^{(2)}l^\mu=\partial X^\mu-k^{-2}
k_\rho \partial X^\rho k^\mu-l^{-2}l_\rho\partial X^\rho l^\mu
\end{equation}
with respect to the scaling symmetry
\begin{equation}
\delta X^\mu=\Lambda^{(1)}(\tau) k^\mu+\Lambda^{(2)}(\tau) l^\mu\, .
\end{equation}
In this way we ensure (local) invariance under the isometric transformations
generated by the two Killing vectors. Using gauge covariant pull-backs it is possible
to eliminate the pull-back of the isometric coordinates, and to reproduce the
isometric couplings in the action in a manifestly covariant way. For example, the
pull-back of the reduced metric in (\ref{IIBwavesBI}), 
${\cal G}_{\mu\nu}\partial X^\mu\partial X^\nu$, can be written as
$g_{\mu\nu}{\cal D}X^\mu{\cal D}X^\nu$, in terms of the covariant pull-backs.
The action is then given by a gauged sigma model of the type considered in
\cite{BJO,BLO, EJL}.

The coupling in (\ref{IIBwavesCS}) plays a key role in the microscopical description
of the $AdS_5\times S^5$ dual giant graviton solution \cite{GMT,HHI}, that we
perform in Section 4. It is due to the fact that we have a second isometric direction
that the $C^{(4)}$ electric potential of this background can couple in the one-dimensional 
worldvolume effective action of the gravitational waves.

The effective action that we will use in the next sections to study the giant graviton
configurations in the $AdS_5\times S^5$ background is then given by: 
\begin{equation}
\label{IIBwavesfull}
S_{\rm W_B} = S^{\rm BI}_{\rm W_B}+S^{\rm CS}_{\rm W_B}
\end{equation}
with $S^{\rm BI}_{\rm W_B}$ given by (\ref{IIBwavesBI}) and $S^{\rm CS}_{\rm W_B}$
by (\ref{IIBwavesCS}).

%%%%%%%%%%%%%%%%%%%%%%%%%%%%%%%%%%%%%%%%%%%%%%%%%%%%%%%%%
\sect{The giant graviton in $AdS_5\times S^5$}

\subsection{The macroscopical description}

The giant graviton solution in $AdS_5\times S^5$ was computed in \cite{GST},
by looking at stable test brane solutions where a D3-brane with
angular momentum in $S^5$ had
expanded to a 3-sphere contained inside the $S^5$. 
We briefly review this construction in order to compare it, in the end, with
our microscopical description. 

Taking the line element for the metric on $AdS_5\times S^5$ as
$ds^2=ds^2_{AdS}+ds^2_{S}$, with
\begin{equation}
ds^2_{AdS}=-(1+\frac{r^2}{L^2})dt^2+\frac{dr^2}{1+\frac{r^2}{L^2}}+
r^2 d\Omega_3^2\, ,
\end{equation}
and
\begin{equation}
ds^2_{S}=L^2(d\theta^2+\cos^2{\theta}d\phi^2+\sin^2{\theta}d\Omega_3^2)\, ,
\end{equation}
the trial solution considered in \cite{GST} has $\theta={\rm constant}$, 
$\phi=\phi(\tau)$, where $t=\tau$ in static gauge, and $r=0$, i.e. it corresponds to
a spherical D3-brane
with radius $L\sin{\theta}$ orbiting the $S^5$ in the $\phi$ direction:
\begin{equation}
\label{testD3}
ds^2=-dt^2+L^2\cos^2{\theta}d\phi^2+L^2\sin^2{\theta}d\Omega_3^2\, .
\end{equation}
This D3-brane carries a non-vanishing magnetic moment with respect to the 
RR 4-form potential of the background, which prevents its collapse to
zero size. Parametrising the unit 3-sphere in (\ref{testD3}) as
\begin{equation}
d\Omega^2_3=d\beta^2_1+\sin^2{\beta_1}(d\beta^2_2+\sin^2{\beta_2}d\beta^2_3),
\end{equation}
the RR 4-form potential is given by
\begin{equation}
C^{(4)}_{\phi\beta_1\beta_2\beta_3}=L^4\sin^4{\theta}\sqrt{g_\beta}\, ,
\end{equation}
where $\sqrt{g_\beta}$ is the volume element on the unit 3-sphere.

Substituting this trial solution into the worldvolume action of the D3-brane,
one arrives at the following Hamiltonian:
\begin{equation}
\label{Hab1}
H=\frac{P_\phi}{L}\sqrt{1+\tan^2{\theta}\Bigl(1-\frac{\tilde N}{P_\phi}
\sin^2{\theta}\Bigr)^2}\, .
\end{equation}
Here ${\tilde N}$ is the integer arising through the quantisation condition
of the 4-form flux on $S^5$
\begin{equation}
\label{quant1}
2\pi^2 T_3=\frac{\tilde N}{L^4}\, ,
\end{equation}
with $T_3$ the tension of the D3-brane, and $P_\phi$ is the angular momentum carried by 
the brane, which is a constant given that $\phi$ is a cyclic coordinate.

The Hamiltonian (\ref{Hab1}) has two stable minima, one for $\sin{\theta}=0$ and another for 
\begin{equation}
\label{giantmac}
\sin^2{\theta}=P_\phi/{\tilde N}\, . 
\end{equation}
The value of the Hamiltonian is in both cases
$E=P_\phi/L$, i.e. both solutions represent massless particles with angular momentum
$P_\phi$. So the first solution corresponds to a pointlike graviton, while the second 
minimum corresponds to a D3-brane with radius $L(P_\phi/{\tilde N})^{1/2}$, which is the
giant graviton solution. The giant graviton  satisfies the stringy exclusion 
principle, since the condition $\sin^2{\theta}\leq 1$ implies an upper bound
on the angular momentum: $P_\phi\leq {\tilde N}$.

%%%%%%%%%%%%%%%%%%%%
\subsection{The microscopical description}

In this section we provide a description of the giant graviton solution in terms
of coincident gravitons expanding into a D3-brane, which is inside the $S^5$-part of the background 
geometry. We will check the correctness of this description by looking whether, for a 
large number of gravitons, it is in agreement with the previous macroscopical description 
in terms of a test D3-brane. 

The similarity between the giant graviton construction in \cite{GST} and the Abelian
description of the dielectric (or magnetic moment) effect for D$p$-branes suggests a 
microscopical description of the giant graviton in terms of gravitons expanding 
into a D3-brane with the topology of a fuzzy 3-sphere. In this description the expansion of 
the gravitons takes place due to their interaction with the RR 4-form potential of the 
background. At the level of the graviton worldvolume effective action this
interaction occurs in the form of a non-Abelian dielectric coupling.

The action that we have proposed in section 2 to describe Type IIB gravitons 
contains two isometric directions, one of which is identified with the direction of
propagation, whereas the other one reflects the fact that the background
on which the gravitons propagate contains a compact direction, which is the
direction of the T-duality transformation involved in the construction. The point now
is to identify these isometries in the background (\ref{testD3}). It is clear that the first 
isometry lays in the $\phi$-direction. The second isometric direction can, on the other
hand, be identified
when one considers the 3-sphere as a $U(1)$-bundle over $S^2$. The natural choice for the 
second isometry is the direction associated to the $U(1)$-fibre. 

Representing the $S^3$ with radius $L\sin{\theta}$ as a submanifold of $\C^2$ with coordinates
$(z_0,z_1)$ satisfying $\bar{z}_0 z_0+\bar{z}_1 z_1 = L^2\sin^2{\theta}$, the Hopf fibering, 
$p:S^3\rightarrow S^2$, is given by a stereographic projection of a point $(z_0, z_1)$ of the 
$S^3$ to a point $z$ of the  $S^2$ (see for example \cite{Nakahara}):
\begin{equation}
p(z_0,z_1)= \left\{ \begin{array}{ll}
                     z=z_1/z_0 & \hsp{.3cm} \mbox{when $z_0\neq 0$} \ ,\\ [.3cm]
                     1/z=z_0/z_1 & \hsp{.3cm} \mbox{when $z_1\neq 0$}\ .
                    \end{array}
            \right.
\end{equation} 
Points on the $S^3$ that differ by an overall factor $\lambda \in U(1)$ get mapped onto the 
same point $z$ of $S^2$: $ p(z_0,z_1) = p(\lambda z_0, \lambda z_1)$. In this way the Hopf map
is dividing out a $U(1)$ fibre in the $S^3$. 
Inversely the coordinates $x_i$ $(i=1,2,3)$ on the $S^2$ with $ x_1^2 + x_2^2 + x_3^2 = R^2$
can be obtained from the Hopf map via 
\be
x_1 = \frac{2}{a}\  \mathfrak{Re} (z_0\ \bar{z_1} ), \hsp{1cm}
x_2 = \frac{2}{a}\  \mathfrak{Im} (z_0\ \bar{z_1} ), \hsp{1cm}
x_3 = \frac{1}{a}\   (| z_0|^2 - |z_1|^2 ),
\ee
where $a$ is an arbitrary parameter with dimension of length, that relates the radius 
$L\sin{\theta}$ of the $S^3$ with the radius $R$ of the $S^2$:
\be
R^2 \ = \ \sum_{i=1}^{3} (x_i)^2  
      \ = \ \frac{1}{a^2} \Bigr[ \bar{z}_0 z_0+\bar{z}_1 z_1 \Bigr]^2
      \ = \ \frac{L^4\sin^4{\theta}}{a^2}.
\ee 
Parametrizing the geometry of the $S^3$ in terms of Euler angles 
\begin{equation}
z_0= L\sin{\theta}\, e^{i(\chi_3+\chi_2)/2}\cos{\tfrac{\chi_1}{2}}\, ,\qquad
z_1= L\sin{\theta}\, e^{i(\chi_3-\chi_2)/2}\sin{\tfrac{\chi_1}{2}}\, ,
\end{equation}
where $\chi_1 \in [0, \pi[$, $\chi_2 \in[0, 2\pi[$ and $\chi_3 \in [0, 4\pi[$,
we get the round metric for $S^3$:
\begin{equation}
ds^2_{S^3}=\frac{L^2\sin^2{\theta}}{4}\Bigl(d\chi_1^2+\sin^2{\chi_1}d\chi_2^2+
(d\chi_3+\cos{\chi_1}d\chi_2)^2\Bigr)\, .
\end{equation}
Here the angles $\chi_1$ and $\chi_2$ parametrise the $S^2$-base manifold and
$\chi_3$ the $S^1$-fibre bundle.  Note that the metric has the necessary twist in the fibre 
in order to obtain the $S^3$ as the global space.
The coordinates $x_i$ on the $S^2$ are then 
given by
\be
x_1 = R \sin \chi_1 \cos \chi_2, \hsp{1cm}
x_2 = R \sin \chi_1 \sin \chi_2, \hsp{1cm}
x_3 = R \cos \chi_1\, ,  
\ee
and we can identify the coordinate $\chi_3$ with the isometric direction in the gravitons 
effective action not associated to the direction of propagation.

Using Euler angles to write the metric of the three-sphere in coordinates adapted to the 
fibre structure, the background metric (\ref{testD3}) and four-form gauge field are given by
\bea
\label{testD3mic}
&& ds^2 = -dt^2+L^2\cos^2{\theta}d\phi^2+\frac{L^2\sin^2{\theta}}{4}
\Bigl(d\chi_1^2+\sin^2{\chi_1}d\chi_2^2+
(d\chi_3+\cos{\chi_1}d\chi_2)^2\Bigr)\, ,\ \nonumber \\  [.3cm]
&& C^{(4)}_{\phi\chi_1\chi_2\chi_3} = -\frac{1}{8} L^4 \sin^4 \theta  \sin \chi_1.
\eea

In order to make a non-Abelian Ansatz, it is convenient to go to Cartesian coordinates 
describing the $S^2$-base manifold of the $S^3$, 
keeping in mind the constraint that $(x_1)^2 + (x_2)^2 + (x_3)^2 = R^2$. In these 
coordinates, the metric and the four-form RR-field become 
\bea
&& ds^2 = -dt^2 + L^2 \cos^2 \theta d\phi^2 
          + \frac{L^2 \sin^2 \theta}{4 R^2} \Bigl[ dx_1^2 + dx_2^2 + dx_3^2 \Bigl] \nn
&& \hsp{2cm}
          + \frac{1}{4} L^2 \sin^2 \theta \Bigl[ 
                  d\chi_3 + \frac{x_3}{R (x_1^2 + x_2^2)} \Bigl( x_1 dx_2 - x_2 dx_1\Bigr) 
               \Bigr]^2 ,
\label{C4} \\[.3cm]
&&C^{(4)}_{\chi_3\phi ij}=\frac{L^4 \sin^4 {\theta}}{8 R^3} \epsilon_{ijk}x^k\, , 
\hsp{1cm} {\rm for}\,\  i,j,k=1,2,3\, .
\nonumber
\eea
We can then make the following non-commutative Ansatz for the 2-sphere that is
parametrised by $x^1,x^2,x^3$:
\begin{equation}
X^i=\frac{R}{\sqrt{N^2-1}}J^i\, ,  
\hsp{2cm}
i=1,2,3
\end{equation}
with the  $J^i$,  forming an $N\times N$ representation
of $SU(2)$ (in our conventions $[J^i,J^j]=2i\epsilon^{ijk}J^k$).
Trivially, with this choice
\begin{equation}
(X^1)^2+(X^2)^2+(X^3)^2= R^2 \unity\, ,
\end{equation}
so we are dealing with a non-commutative version of the $S^2$ contained in 
(\ref{testD3mic}). Therefore, with this non-commutative Ansatz, the 3-sphere 
becomes an $S^1$-bundle over a fuzzy $S^2$.  This situation is forced by the topology 
of the space in which the Type~IIB gravitons propagate, having  an extra compact 
$S^1$ direction. Thus, the physical picture will correspond to the gravitons expanding into 
a non-commutative D3-brane, described in coordinates that reflect a 
$S^2_{\rm fuzzy}\times S^1$ structure. To see that 
this is the right microscopical picture we have to check that in the limit in which the 
number of gravitons is large we recover the description in \cite{GST}, in terms of a 
D3-brane with the topology of a classical, commutative, $S^3$. We will carry out 
this calculation in the remaining part of this section.

The action (\ref{IIBwavesfull}) for Type IIB waves in the $AdS_5 \x S^5$ background
described by (the non-commutative version of) the coordinates (\ref{C4}), contains
\begin{eqnarray}
&& E_{00}=-1,  \qquad E_{0i}=0,\qquad 
E_{ij}=\frac{L^2 \sin^2\theta}{4 R^2} \Bigl[ \delta_{ij}
                - \frac{\sin\theta}{R\cos{\theta}}\epsilon_{ijk}X^k \Bigr] \nn
&& Q^i{}_j=\delta^i_j \ -\ \frac{L^4\sin^3{\theta}\cos{\theta}}{4R\sqrt{N^2-1}}
                              \epsilon^{ijk}  X_k
\ +\  \frac{L^4\sin^4{\theta}}{4 R^2\sqrt{N^2-1}}(X^iX_j-\delta^i_j X^2)
\end{eqnarray}
in the Born-Infeld part, while the contributions to the Chern-Simons part vanish.
We then find:
\begin{equation}
S_{\rm W_B}=-T_0 \int d\tau\,  {\rm STr}\Bigl\{ \frac{1}{L\cos{\theta}}
\sqrt{\unity - \frac{L^4\sin^4{\theta}}{2 R^2\sqrt{N^2-1}} X^2
             + \frac{L^8\sin^6{\theta}\cos^2{\theta}}{16 R^2(N^2-1)}X^2
             + \frac{L^8\sin^8{\theta}}{16 R^4(N^2-1)}X^2X^2}    \Bigr\}\, ,
\end{equation}
where we have dropped those contributions to ${\rm det}\ Q$ that will vanish
when taking the symmetrised average involved in the symmetrised trace prescription.
In our description, since the direction of propagation is isometric, we are effectively dealing 
with a static configuration, for which we can compute the potential as minus the Lagrangian.
In the limit $L^4\sin^4{\theta}\ll \sqrt{N^2-1}$ we can approximate the square root by
\begin{equation}
\unity-\frac{L^4\sin^4{\theta}}{4R^2\sqrt{N^2-1}}X^2+\frac{L^8\sin^6{\theta}\cos^2{\theta}}
{32R^2(N^2-1)}X^2\, ,
\end{equation}
and since $X^2=R^2\unity$, we have for the potential
\begin{equation}
V_{\rm W_B}(\theta)=\frac{NT_0}{L\cos{\theta}}\Bigl(1-\frac{L^4\sin^4{\theta}}
{4\sqrt{N^2-1}}+\frac{L^8\sin^6{\theta}\cos^2{\theta}}{32(N^2-1)}\Bigr)\, ,
\end{equation}
which can be seen as the first order expansion of
\begin{equation}
V_{\rm W_B}(\theta)=\frac{NT_0}{L\cos{\theta}}\sqrt{1-\frac{L^4\sin^4{\theta}}
{2\sqrt{N^2-1}}+\frac{L^8\sin^6{\theta}}{16(N^2-1)}}
\end{equation}
in the same limit above. Introducing $\cos{\theta}$ inside the
square root we have
\begin{equation}
\label{potmic}
V_{\rm W_B}(\theta)=\frac{NT_0}{L}\sqrt{1+\tan^2{\theta}
\Bigl(1-\frac{L^4\sin^2{\theta}}{4\sqrt{N^2-1}}\Bigr)^2}\, .
\end{equation}
This potential has two minima, the point-like graviton at $\sin{\theta}=0$,
and a solution that should correspond to the giant graviton solution at 
\begin{equation}
\label{giantmic}
\sin^2{\theta}=\frac{4\sqrt{N^2-1}}{L^4}\, .
\end{equation} 
Both solutions have an energy $E=NT_0/L$, and are therefore associated to massless particles
with angular momentum $P_\phi=NT_0$. There is an upper bound on the angular momentum that comes 
from the condition $\sin^2{\theta}\leq 1$, which implies that
\begin{equation}
P_\phi=NT_0\leq \frac{T_0}{4}\sqrt{L^8+16}\, .
\label{bound}
\end{equation}
Comparing the potential (\ref{potmic}), the radius (\ref{giantmic}) and the upper bound
(\ref{bound}) with their Abelian counterparts in the macroscopical derivation,  
we find that there is exact agreement in the large $N$ limit when
\begin{equation}
\label{condi}
\frac{{\tilde N}}{P_\phi}=\frac{L^4}{4N} \qquad \Longleftrightarrow
\qquad P_\phi = 8 \pi^2 N T_3\, .
\end{equation}
Making use of the fact that in the microscopical description the total angular 
momentum is quantised in terms of the tension of the gravitational waves, $P_\phi=NT_0$,
we obtain the following relation between the tension $T_0$ of the waves and the tension $T_3$
of the D3-brane: 
\be
T_0=8\pi^2 T_3,
\ee 
which is indeed satisfied, given that the isometric transverse
direction is an angular variable with period $4\pi$.

%%%%%%%%%%%%%%%%%%%%%%%%%%%%%%%%%%%%%%%%%%%%%%%%%%
\sect{The dual giant graviton in $AdS_5\times S^5$}

\subsection{The macroscopical description}

The dual giant graviton solution in $AdS_5\times S^5$ was computed in
\cite{GMT,HHI}, by looking at stable test brane solutions where the
D3-brane with angular momentum in $S^5$ expands to the 3-sphere
contained inside the $AdS_5$ component of the spacetime.
For this giant graviton type of
solution there is no upper bound for the angular momentum, because
it expands in the non-compact part of the geometry. Let us briefly summarise
the construction in these references in order to compare it, in the end, 
with our microscopical description.

The trial solution in this case is taken with $r={\rm constant}$, $\phi=\phi(\tau)$
and $\theta=0$, which corresponds to a spherical D3-brane with radius $r$
orbiting the $S^5$ in the $\phi$ direction:
\begin{equation}
\label{D3testdual}
ds^2=-(1+\frac{r^2}{L^2})dt^2+r^2d\Omega_3^2+L^2d\phi^2\, .
\end{equation}
This D3-brane carries a non-vanishing dipole moment with respect to the
RR 4-form potential, which prevents its collapse to zero size.
Parametrising the unit 3-sphere in (\ref{D3testdual}) as
\begin{equation}
d\Omega_3^2=d\alpha_1^2+\sin^2{\alpha_1}(d\alpha_2^2+\sin^2{\alpha_2}
d\alpha_3^2)\, ,
\end{equation}
we have
\begin{equation}
C^{(4)}_{0\alpha_1\alpha_2\alpha_3}=-\frac{r^4}{L}\sqrt{g_\alpha}\, .
\end{equation}
Substituting this trial solution into the worldvolume action of the D3-brane,
one arrives at the following Hamiltonian \cite{GMT,HHI}:
\begin{equation}
\label{Hmacdual}
H=\frac{1}{L}\left[\sqrt{\Bigl(1+\frac{r^2}{L^2}\Bigr)
                         \Bigl(P_\phi^2+\frac{{\tilde N}^2r^6}{L^6}\Bigr)}
-\frac{{\tilde N}r^4}{L^4}\right]\, ,
\end{equation}
where ${\tilde N}$ is given by (\ref{quant1}).

The stable solutions correspond to $r=0$, the point-like graviton, and  
$r=L\sqrt{P_\phi/{\tilde N}}$, the dual giant graviton, both of which have
energy $E=P_\phi/L$, representing massless particles with angular momentum
$P_\phi$. Contrary to the giant graviton solution of the previous section,
the dual giant graviton solution does not satisfy the stringy exclusion
principle, because the absence of an upper bound for its radius implies that
$P_\phi$ is neither bounded.

%%%%%%%%%%%%%%%%%%%%%%%%%%%%%%%%%%%%%%%
\subsection{The microscopical description}

We now want to provide a microscopical description of the
dual giant graviton solution. This description will be in terms of
coincident gravitons expanding into a D3-brane which is now expanded
in a fuzzy surface contained inside the $AdS_5$ part of the geometry.
As in the previous section, we represent the $S^3$ in (\ref{D3testdual})
as an $S^1$ bundle over $S^2$, take Euler angles such that
\begin{equation}
\label{roundS3dual}
ds_{S^3}^2=\frac{r^2}{4}\Bigl(d\chi_1^2+\sin^2{\chi_1}d\chi_2^2+
(d\chi_3+\cos{\chi_1}d\chi_2)^2\Bigr)
\end{equation}
and identify the coordinate $\chi_3$ parametrising the $S^1$ with the compact 
isometric direction coming from the T-duality construction. The isometry 
associated to the propagation direction is again taken to be $\phi$.
The background metric (\ref{D3testdual}) is then given by
\begin{equation}
\label{D3testdualan}
ds^2=- (1+\frac{r^2}{L^2}) dt^2+\frac{r^2}{4}\Bigl(d\chi_1^2+\sin^2{\chi_1}d\chi_2^2+
(d\chi_3+\cos{\chi_1}d\chi_2)^2\Bigr)+L^2d\phi^2\, .
\end{equation}
Taking Cartesian coordinates in the $S^2$ that is parametrised by
$\chi_1$ and $\chi_2$
\begin{eqnarray}
\label{2spheredual}
x^1 = R\sin{\chi_1}\cos{\chi_2}, \hsp{1,3cm}
x^2 = R\sin{\chi_1}\sin{\chi_2}, \hsp{1,3cm}
x^3 = R\cos{\chi_1}\, ,
\end{eqnarray}
the metric and 4-form potential take the form
\bea
\label{C3elec}
&& ds^2 =  - (1+\frac{r^2}{L^2}) dt^2 + L^2 d\phi^2  
            + \frac{r^2}{4R^2} \Bigl(dx_1^2 + dx_2^2 + dx_3^2\Bigr)\nn
&& ~\hsp{1.4cm} 
            + \frac{r^2}{4} \Bigl[ d\chi_3 
               +\frac{x_3}{R(x_1^2 + x_2^2)}\Bigl( x_1 dx_2 - x_2dx_1\Bigr) \Bigr]^2
\nn
&& C^{(4)}_{\chi_3 0ij}=-\frac{r^4}{8R^3L}\epsilon_{ijk}x^k
\eea
for $i,j,k=1,2,3$. Thus, the natural non-commutative Ansatz to make in this case is
\begin{equation}
X^i=\frac{R}{\sqrt{N^2-1}}J^i\, ,
\end{equation}
with $J^i$ forming an $N\times N$ representation of $SU(2)$. Our
description of the dual giant graviton will again be in terms of gravitons
expanding into a non-commutative D3-brane with topology
$S^2_{\rm fuzzy}\times S^1$, the validity of which should be tested by 
checking the agreement with the macroscopical description for large number
of gravitons.

The action (\ref{IIBwavesfull}) for Type IIB waves in the particular background
defined by (the non-commutative version of) (\ref{C3elec})
contains:
\begin{eqnarray}
&&E_{00}=-(1+\frac{r^2}{L^2})\, ,\qquad E_{0i}=0\, ,\qquad
E_{ij}=\frac{r^2}{4R^2}\delta_{ij}\, , \\
&&Q^i_j=\delta^i_j-\frac{Lr^3}{4R\sqrt{N^2-1}}\epsilon^{ijk}X^k\, .
\end{eqnarray}
The Born-Infeld part of the action then takes the form:
\begin{equation}
S^{\rm BI}_{\rm W_B}=-\frac{T_0}{L}\int d\tau\, {\rm STr}\Bigl\{\sqrt{\Bigl(1+\frac{r^2}{L^2}\Bigr)
\Bigl(\unity+\frac{L^2 r^6}{16R^2(N^2-1)}X^2\Bigr)}\Bigr\}\, ,
\end{equation}
while the Chern-Simons part is, in turn, given by:
\begin{equation}
S^{\rm CS}_{\rm W_B}=-T_0\int d\tau\, {\rm STr}\Bigl\{ iP[(\incl_X\incl_X)\incl_l C^{(4)}]\Bigr\}=
\int d\tau\, \frac{NT_0}{4\sqrt{N^2-1}}\frac{r^4}{L}\, .
\end{equation}
We then have a potential:
\begin{equation}
V_{\rm W_B}(r)=\frac{T_0}{L}\,{\rm STr}\, \sqrt{\Bigl(1+\frac{r^2}{L^2}\Bigr)
\Bigl(\unity+\frac{L^2r^6}{16R^2(N^2-1)}X^2\Bigr)}-\frac{NT_0}{4\sqrt{N^2-1}}
\frac{r^4}{L}\, .
\end{equation}
In the limit $Lr^3\ll \sqrt{N^2-1}$ we can approximate the square root by
its first order expansion, take the symmetrised trace (which to this order
will only produce a factor of $N$ in front of the action) and regard the
remaining expression as the first order expansion of the potential
\begin{equation}
\label{potmicdual}
V_{\rm W_B}(r)=\frac{NT_0}{L} \sqrt{\Bigl(1+\frac{r^2}{L^2}\Bigr)
\Bigl(1+\frac{L^2r^6}{16(N^2-1)}\Bigr)}-\frac{NT_0}{4\sqrt{N^2-1}}
\frac{r^4}{L}\, ,
\end{equation}
exactly as we did in the previous section.

This potential has two minima, at $r=0$, corresponding to the point-like
graviton, and at
\begin{equation}
r^2=\frac{4\sqrt{N^2-1}}{L^2}\, ,
\end{equation}
which should correspond to the dual giant graviton solution.
Both minima have energy $E=NT_0/L$. Comparing these results with the
Hamiltonian and the radius of the macroscopical derivation, we find that,
in the large $N$ limit, there is exact agreement when the condition (\ref{condi})
is fullfilled.

%%%%%%%%%%%%%%%%%%%%%%%%%%%%%%%%%%%%%%%%%%%%%%%%
\section{Discussion}

In this paper we have discussed a explicit Matrix action which is solved
by a non-commutative 3-sphere. This Matrix model arises as an action for coincident Type IIB
gravitational waves. This action is constructed using T-duality from the action for
coincident Type IIA gravitational waves of \cite{JL1,JL2}, and is such that the T-duality
direction appears as a special isometric direction, on which neither the background fields
nor the pull-backs depend. The presence of this compact isometric direction suggests the
representation of the fuzzy 3-sphere solution as an $S^1$ bundle over a fuzzy 2-sphere base
manifold. Accordingly, our solution does not show manifest $SO(4)$ covariance, 
this invariance being broken down to $SU(2)\times U(1)$. The $SO(4)$
invariance should, still, be present in a non-manifest way, in the same fashion than the
$SO(4)$ covariance of the classical 3-sphere is not explicit when the $S^3$ is described 
as an $S^1$ Hopf fibering. This fuzzy 3-sphere 
should occur as a BPS solution of the Matrix model, preserving the same 
supersymmetries as the point-like graviton. We have not checked out this property explicitly,
but it is expected from the agreement with the dual macroscopical
description, in which the supersymmetry properties of these spherical
configurations are demonstrated explicitly \cite{GMT,HHI}. 
 
Odd non-commutative spheres have been previously discussed in the
literature \cite{GR,R1,R2}. We would like to mention some relations between these different
constructions.

In references \cite{GR,R1,R2} an $SO(4)$-covariant matrix realisation of the condition 
$\sum_{i=1}^{4}X_i^2=\unity$ is found, in terms of matrices acting on some vector space.
As for any fuzzy sphere with more than two dimensions,
this matrix algebra contains more representations than is necessary to describe
functions on the 3-sphere, so certain projections need to be done to eliminate the excess
of degrees of freedom. The non-commutative 3-sphere that results has manifest $SO(4)$ covariance,
but its non-commutativity cannot be removed completely in the large $N$ limit. On the contrary,
the 3-sphere solution that we have constructed in this paper
shows only manifest $SU(2)\times U(1)$ covariance, but approaches neatly 
the classical $S^3$ in the large $N$ limit, where all the non-commutativity disappears. 
The fuzzy 3-sphere of \cite{GR,R1,R2} can be represented in the large $N$ limit
as a fibration of a 2-sphere over an interval\footnote{We thank Sanjaye Ramgoolam for pointing this
out to us and for the discussion below for finite $N$.}.
For finite $N$, it is expected that this representation is in terms of
a discrete $S^1$ fibration
over a 2-sphere, where both the $S^2$ and the $S^1$ are non-commutative. The reason for this is
that the algebra of functions on $S^1$ is infinite dimensional so it cannot fit 
inside a finite Matrix Algebra as that of the fuzzy $S^3$. 
It would be interesting to develop a careful description of this $S^1$ fibration structure.
 
The fuzzy $S^3$ solution with manifest $SO(4)$ covariance 
is expected to play a role in a plausible description of Type IIB gravitons in terms of
an action with no isometry directions (other than the direction of propagation). We should
mention at this point that the reason why we have not succeeded in constructing such an action
is a technical one, namely the impossibility of restoring the dependence on the T-duality direction 
in the non-Abelian case. Should the construction of such an action be possible, an $SO(4)$
covariant solution as that of \cite{GR,R1,R2} would most likely be the right Ansatz for the study
of giant graviton configurations.

In fact, an explicit physical realisation of the fuzzy 3-sphere discussed above is in the context of
non-BPS Type IIB D0-branes \cite{GR}. The 3-sphere arises as a solution of the action for
coincident non-BPS Type IIB D0-branes \cite{JM,MS}
in a flat background with constant 5-form field strength.
In this construction the
tachyonic field of the D0-branes plays an essential role in describing the fuzzy $S^3$ as a
subspace of a fuzzy $S^4$. This solution cannot however be interpreted in the large $N$ limit 
as a spherical D3-brane, because its dipole moment vanishes for large 
$N$ \footnote{As discussed in \cite{GR} this could be due to 
the fact that the background they consider is not
a consistent supergravity background.}. Our 3-sphere solution, on the other hand, 
has an interpretation as
a spherical D3-brane, since its dipole, or magnetic moment, coupling to the
5-form field strength resembles that of a spherical D3-brane, not only at finite $N$ but also in
the large $N$ limit\footnote{This result seems to confirm the
previous observation in \cite{GR}, since our solution occurs in a consistent supergravity 
background. One should take into account however that we are
dealing with gravitational waves and not with non-BPS
D0-branes.}. 
A dual macroscopical description in terms of a single
D3-brane with whom to compare the microscopical description in \cite{GR}
is however not possible, given that it is not known how to dissolve non-BPS D0-branes 
in the worldvolume of a BPS D3-brane\footnote{It is likely 
that considering a D3, anti-D3 system would help for this purpose.}.
Our solution on the contrary arises when studying the polarisation of,
BPS, gravitational waves. For this system not only the Lagrangian is defined in a 
better way, since there is no uncontrolled tachyon dynamics, but, 
moreover, a dual description in terms of a single D3-brane
exists, which can be compared in the large $N$ limit with the
microscopical description.
The agreement between the two approaches provides in fact the strongest support
to our construction.

%%%%%%%%%%%%%%%%%%%%%%%%%%%%%%%%%%%%%%%5
\vspace{1cm}
\noindent
{\bf \large Acknowledgements}\\
The authors are grateful to Jos Gheerardyn, Walter Troost and especially Sanjaye Ramgoolam for very
useful discussions.  The work of B.J.~has been done as a post-doctoral fellow of the
F.W.O.-Vlaanderen. B.J.~was also partially supported by the F.W.O.-project G0193.00N, by 
the European Commission R.T.N.-program HPRN-CT-2000-00131 and by the Belgian Federal Office for
Scientific, Technical and Cultural Affairs through the Interuniversity Attraction Pole
P5/27.  The work of Y.L. has been partially supported by CICYT grant BFM2000-0357
(Spain). D.R-G. was supported in part by a F.P.U. Fellowship from M.E.C. (Spain).
He would like to thank the Departamento de F\'{\i}sica Te\'orica at
Universidad Aut\'onoma de Madrid for its hospitality while part of this work was done.

%%%%%%%%%%%%%%%%%%%%%%%%%%%%%%%%%%%%%%%%%%%%%
%%%%%%%%%%%%%%%%%%%%%%%%%%%%%%%%%%%%%%%%%%%%%%%

%%%%%%%%%%%%%%%%%%%%%%%%%%%%%%%%%%%%%%%%%%%%%%%%%%%%%%%%%
%\newpage

%%%%%%%%%%%%%%%%%%%%%%%%%%%%%%%%%%%%%%%%%%%%%%%%%%%%%%%%%
\end{document}